\documentclass{llncs}
\usepackage{}
\usepackage{amsmath}
\usepackage{cases}
\usepackage{amsfonts}
\usepackage{amssymb}
\usepackage{makeidx}  % allows for indexgeneration
\usepackage{amsfonts,amssymb}
\usepackage{graphicx}
\begin{document}

\title{Refuting Unique Game Conjecture}
%\subtitle{[extended abstract]}
%
\titlerunning{Refuting}  % abbreviated title (for running head)
\renewcommand\thefootnote{}

\author{Peng Cui}

\authorrunning{Peng Cui}   % abbreviated author list (for running head)
%
%%%% modified list of authors for the TOC (add the affiliations)
\tocauthor{Peng Cui}
\institute{Key Laboratory of Data Engineering and Knowledge Engineering, MOE,
School of Information Resource Management, Renmin University of China, Beijing
100872, P. R. China.\\
\email{cuipeng@ruc.edu.cn}}

\maketitle              % typeset the title of the contribution

\begin{abstract}
In this short note, the author shows that the gap problem of some $k$-CSPs with the support of its predicate the ground of a balanced pairwise independent distribution can be solved by a modified version of Hast's Algorithm BiLin that calls Charikar\&Wirth's SDP algorithm for two rounds in polynomial time, when $k$ is sufficiently large, the support of its predicate is combined by the grounds of three biased homogeneous distributions and the three biases satisfy certain conditions. To conclude, the author refutes Unique Game Conjecture, assuming $P\ne NP$.
\end{abstract}

\section{Introduction}
Max $k$-CSP is the task of satisfying the maximum fraction of constraints when each constraint involves $k$ variables, and each constraint accepts the same collection $C\subseteq G^k$ of local assignments. A challenging question is to identify constraint satisfaction problems (CSPs) that are extremely hard to approximate, so much so that they are NP-hard to approximate better than just outputting a random assignment. Such CSPs are called approximation resistant; famous examples include Max 3-SAT and Max 3-XOR\cite{h}. A lot is known about such CSPs of arity at most four\cite{ha}, but for CSPs of higher arity, results have been scattered.

To make progress, conditional results are obtained assuming the Unique Game Conjecture (UGC) of \cite{k}. Under UGC, \cite{am} shows that a CSP is approximation resistant if the support of its predicate is the ground of a balanced pairwise independent distribution. Traditionally, negations are allowed free in definition of $k$-CSP. In \cite{ah}, the authors investigate $k$-CSP with no negations of variables and prove such $k$-CSP with the support of its predicate the ground of a biased pairwise independent distribution or uniformly positively correlated distribution or is approximation resistant in biased sense under Unique Game Conjecture.

The following is Theorem 3.1 in \cite{am}.

\begin{theorem}
Let $k\ge 3$ be an integer, and $C$ be a subset of $G^k$ and the ground of a balanced pairwise independent distribution. For arbitrarily small constant $\varepsilon$, it is UG-hard to distinguish the following two cases given a instance $P$ of Max $C$e:

\begin{itemize}
\item Completeness: $\mathrm{val}(P)\ge 1-\varepsilon$.
\item Soundness: $\mathrm{val}(P)\le\frac{|C|}{2^k}+\varepsilon$.
\end{itemize}

\end{theorem}

In this short note, the author shows that the gap problem of this type of $k$-CSP can be solved by a modified version of Hast's Algorithm BiLin in polynomial time that calls Charikar\&Wirth's SDP algorithm\cite{cw} for two rounds, when $k$ is sufficiently large, the support of its predicate is combined by the grounds of three biased homogeneous distributions and the three biases satisfy certain conditions. To conclude, the author refutes Unique Game Conjecture, assuming $P\ne NP$.

\begin{theorem}
Unique Game Conjecture does not hold true, assuming $P\ne NP$.
\end{theorem}

This work has an origin that conditionally strengthens the previous known hardness for approximating Min 2-Lin-2 and Min Bisection, assuming a claim that refuting Unbalanced Max 3-XOR under biased assignments is hard on average\cite{c}. In this paper, the author defines "bias" to be a parameter of pairwise independent subset (distribution), while he defines "bias" to be the fraction of variables assigned to value 1 in \cite{c}. The author notices that biased pairwise independent distribution is defined in \cite{am,ah} and uniformly positively correlated distribution is defined in \cite{ah}.

\section{Definitions}
As usual, let $[q]=\{1,2,\cdots,q\}$, and $-[q]=\{-q,-q+1,\cdots,-1\}$.

Let $G=\{1,-1\}$, here 1 represent "0/false" and -1 represent "1/true" in standard Boolean algebra.

Random variables are denoted by italic boldface letters, such as $\vec z$. Suppose $\varphi$ is a distribution over $G^k$, the ground of $\varphi$ is defined as

$$G_\varphi=\{\varphi(\vec z)>0|\vec z\in G^k\}.$$

\begin{definition}
For some $0<\gamma<1$, a distribution $\varphi$ over $G^k$ is biased pairwise independent if for every coordinate $i\in[k]$,

$$\mathbb{P}[\vec z_i=1]=\gamma$$

\noindent and for every two distinct coordinates $i,j\in[k]$,

$$\mathbb{P}[\vec z_i=1,\vec z_j=1]=\gamma^2,$$

\noindent where $\vec z$ is a random element drawn from $G^k$ by $\varphi$. $\gamma$ is called bias of $\varphi$. If $\gamma=\frac{1}{2}$, we say $\varphi$ is balanced pairwise independent.
\end{definition}

The author notices the fact that a distribution over $G^k$ can be thought as a linear superposition of several distributions over $G^k$.

\begin{definition}
Given $m$ distributions $\varphi_l$ over $G^k$ with disjoint grounds $G_{\varphi_l}$, let $\psi$ is a distribution over $[m]$ with $\psi_l>0$ for each $l\in[m]$, and $\varphi$ be the distribution over $G^k$ such that

$$\varphi(\vec z)=\sum^{m}_{l=1}\psi_l\varphi_l(\vec z),$$

\noindent for each $\vec z\in G^k$. We say $\varphi_l$'s are disguised by $\psi$ to $\varphi$.
\end{definition}

\section{Proof of Theorem 2}
We construct the instances by three homogeneous distributions that are uniformly negatively correlated.

For each $l\in[3]$, suppose $\gamma_l$ is a constant satisfying $0\le\gamma_l\le 1$ and $\gamma_l k$ is an integer and $\rho_l$'s are constants such that

$$\gamma_1=\textstyle\frac{1}{2}+\rho_1\textstyle\frac{1}{\sqrt{k}},\gamma_2=\textstyle\frac{1}{2}-\rho_2\textstyle\frac{1}{\sqrt{k}},
\gamma_3=\textstyle\frac{1}{2}+\rho_3\textstyle\frac{1}{\sqrt{k}}.$$

Let $G_m$ denote the subset of $G^k$ including all $k$-tuples with exactly $m$ $1$. For each $l\in[3]$, let $\varphi_l$ be the uniform distribution over $G_{\gamma_l k}$, called {\it $\gamma_l$-biased homogeneous distribution}, which satisfies the following two properties:

For every coordinate $i\in[k]$,

$$\mathbb{P}[\vec z_i=1]=\textstyle\frac{1}{2},$$

\noindent and for every two distinct coordinates $i,j\in[k]$,

$$\mathbb{P}[\vec z_i=1,\vec z_j=1]=\textstyle\frac{k}{k-1}\gamma_l^2-\textstyle\frac{1}{k-1}\gamma_l,$$

\noindent where $\vec z$ is a random element drawn by $\varphi_l'$.

We can prove the following lemma. (Section 4)

\begin{lemma}
Suppose $C=G_{\varphi_1}\cup G_{\varphi_2}\cup G_{\varphi_3}$, and $P^{(3)}(y)$ is the tri-linear term of the Fourier spectra of $C$. When $k$ is sufficiently large, there are three absolute constants $\rho_1$, $\rho_2$ and $\rho_3$ satisfying $0<\rho_1<\rho_2\approx\rho_3$, $\gamma_l k$'s are integers, and a distribution $\psi$ over $[3]$ such that:

\begin{enumerate}
\item $\varphi_l$'s are disguised by $\psi$ to a balanced pairwise independent distribution $\varphi$.
\item $P^{(3)}(y)\ge\iota$ for any $y\in C$, where $\iota=\Omega(\frac{1}{\sqrt{k}})$.
\end{enumerate}

\end{lemma}

In the dictatorship test, for every three distinct coordinates $i_1,i_2,i_3\in[k]$, let

$$\mathbb{P}[\vec z_{i_1}=1,\vec z_{i_2}=1,\vec z_{i_3}=1]\triangleq a,$$

\begin{equation*}
\begin{split}
&\mathbb{P}[\vec z_{i_1}=1,\vec z_{i_2}=1,\vec z_{i_3}=-1]=\mathbb{P}[\vec z_{i_1}=1,\vec z_{i_2}=-1,\vec z_{i_3}=1] \\
&=\mathbb{P}[\vec z_{i_1}=1,\vec z_{i_2}=1,\vec z_{i_3}=-1]\triangleq b,
\end{split}
\end{equation*}

\begin{equation*}
\begin{split}
&\mathbb{P}[\vec z_{i_1}=1,\vec z_{i_2}=-1,\vec z_{i_3}=-1]=\mathbb{P}[\vec z_{i_1}=-1,\vec z_{i_2}=1,\vec z_{i_3}=-1] \\
&=\mathbb{P}[\vec z_{i_1}=-1,\vec z_{i_2}=-1,\vec z_{i_3}=1]\triangleq c,
\end{split}
\end{equation*}

$$\mathbb{P}[\vec z_{i_1}=-1,\vec z_{i_2}=-1,\vec z_{i_3}=-1]\triangleq d,$$

\noindent where $\vec z$ is a random element drawn by $\varphi$.

Then

\begin{equation*}
\begin{pmatrix}
 1 & 3 & 3 & 1 \\
 1 & -3 & 3 & -1 \\
 1 & 1 & 0 & 0 \\
 0 & 0 & 1 & 1
\end{pmatrix}
\begin{pmatrix}
 a \\
 b \\
 c \\
 d
\end{pmatrix}
=
\begin{pmatrix}
1 \\
\alpha \\
\frac{1}{4} \\
\frac{1}{4}
\end{pmatrix},
\end{equation*}

\noindent where

$$\textstyle\alpha\approx 8\rho_1(\frac{1}{4}-\rho_2^2)\textstyle\frac{1}{k\sqrt{k}},$$

\noindent which reduces to

\begin{equation*}
\begin{pmatrix}
 a \\
 b \\
 c \\
 d
\end{pmatrix}
=
\begin{pmatrix}
 \frac{1}{8}+\frac{\alpha}{8}\\
 \frac{1}{8}-\frac{\alpha}{8}\\
 \frac{1}{8}+\frac{\alpha}{8}\\
 \frac{1}{8}-\frac{\alpha}{8}
\end{pmatrix}.
\end{equation*}

By Theorem 1, given an instance $P$ as in the statement of Lemma 1, for arbitrarily small constant $\varepsilon$, it is UG-hard to distinguish the following two cases: $\mathrm{val}(P)\ge 1-\varepsilon$; $\mathrm{val}(P)\le\frac{|C|}{2^k}+\varepsilon$.

On the other hand, consider the Fourier spectra of $P$, since $C$ is folded and $\vec z$ is balanced pairwise independent, there is no linear or bi-linear term. Let $I_{123}^{(3)}$ be the sum of tri-linear terms containing three variables with coordinate $1$, $2$ and $3$ respectively in the Fourier spectra of $P$. Suppose $\mathrm{val}(P)\ge 1-\varepsilon$ for some $\varepsilon$, there is an assignment $f^*$ under which $I_{123}^{(3)}\ge\Omega(\frac{1}{k^2\sqrt{k}})$ (cf. Lemma 4 in \cite{ha}).

Let $I_{123}^{(2)}$ be the sum of bi-linear terms defined as: For each tri-linear term $a_{i_1i_2i_3}x^{(1)}_{i_1}x^{(2)}_{i_2}x^{(3)}_{i_3}$ in $I_{123}^{(3)}$, introduce a bi-linear term $a_{i_1i_2i_3}x^{(1)}_{i_1}x_{i_2i_3}$, where $x^{(23)}_{i_2i_3}$'s are new variables in $G$, where $i_1\in[M]$ and $i_2,i_3\in[N]$.

We modify Hast's Algorithm BiLin as follows:

\begin{itemize}
\item Step 1, run Charikar\&Wirth's SDP algorithm for the first round on $I_{123}^{(2)}$ to get an assignment $f^{(1)}$ on $x^{(1)}_{i_1}$'s and $x^{(23)}_{i_2i_3}$'s.
\item Step 2, run Charikar\&Wirth's SDP algorithm for the second round on $I^{(3)}$ subject to $f^{(1)}$ to get an assignment $f^{(2)}$ to $x^{(2)}_{i_2}$'s and $x^{(3)}_{i_3}$'s.
\item Step 3, let $f:=f^{(1)}$ for $x^{(1)}_{i_1}$'s and let $f:=f^{(2)}$ for $x^{(2)}_{i_2}$'s and $x^{(3)}_{i_3}$'s.
\item Step 4, same as Step 3 in the original algorithm.
\item Step 5, same as Step 4 in the original algorithm.
\end{itemize}

The first round returns $f^{(1)}$ under which $I_{123}^{(2)}$ is at least $\Omega(\frac{1}{k^2\sqrt{k}\log k})$ (cf. Lemma 5 in \cite{cw}). By enumeration arguments, there is an assignment $f'$ to $x^{(2)}_{i_2}$'s and $x^{(3)}_{i_3}$'s under which $I^{(3)}$ subject to $f^{(1)}$ is at least $\Omega(\frac{1}{k^2\sqrt{k}\log k})$. Hence the second round returns $f^{(2)}$ under which $I^{(3)}$ subject to $f^{(1)}$ is at least $\Omega(\frac{1}{k^2\sqrt{k}\log^2 k})$ (cf. Lemma 5 in \cite{cw}).

Therefore, the modified version of BiLin returns a solution of $P$ with value at least $\frac{|C|}{2^k}+\kappa$, where

$$\textstyle\kappa=\Omega(\frac{1}{k^2\sqrt{k}\log^2kk})^3=\Omega(\frac{1}{\mathrm{poly}(k)}),$$

\noindent (cf. Theorem 3 in \cite{ha}).

Therefore, Unique Game Conjecture does not hold true, assuming $P\ne NP$. The proof of Theorem 2 is accomplished.

\section{Proof of Lemma 1}
Let $m=3$, consider the linear equations with $\psi_l$ for $l\in[m]$,

\begin{equation*}
\begin{pmatrix}
 1 & 1 & 1 \\
 \gamma_1 & \gamma_2 & \gamma_3 \\
 \frac{k}{k-1}\gamma_1^2-\frac{1}{k-1}\gamma_1 &
 \frac{k}{k-1}\gamma_2^2-\frac{1}{k-1}\gamma_2 &
 \frac{k}{k-1}\gamma_3^2-\frac{1}{k-1}\gamma_3
\end{pmatrix}
\begin{pmatrix}
\psi_1 \\
\psi_2 \\
\psi_3
\end{pmatrix}
=
\begin{pmatrix}
1 \\
\textstyle\frac{1}{2} \\
\textstyle\frac{1}{4}
\end{pmatrix},
\end{equation*}

\noindent which reduces to

\begin{equation*}
\begin{pmatrix}
 1 & 1 & 1 \\
 \rho_1 & -\rho_2 & \rho_3 \\
 \rho_1^2 & \rho_2^2 & \rho_3^2 &
\end{pmatrix}
\begin{pmatrix}
\psi_1 \\
\psi_2 \\
\psi_3
\end{pmatrix}
=
\begin{pmatrix}
1 \\
0 \\
\textstyle\frac{1}{4}
\end{pmatrix},
\end{equation*}

\noindent or

\begin{equation*}
\begin{pmatrix}
\psi_1 \\
\psi_2 \\
\psi_3
\end{pmatrix}
=
\begin{pmatrix}
 \frac{-\rho_2\rho_3+\textstyle\frac{1}{4}}{(\rho_1+\rho_2)(\rho_1-\rho_3)} \\
 \frac{\rho_3\rho_1+\textstyle\frac{1}{4}}{(\rho_2+\rho_3)(\rho_2+\rho_1)} \\
 \frac{-\rho_1\rho_2+\textstyle\frac{1}{4}}{(\rho_3-\rho_1)(\rho_3+\rho_2)}
\end{pmatrix}.
\end{equation*}

For $0<\rho_1<\rho_3$ and $0<\rho_2$, $\psi_l>0$ if and only if $\rho_1\rho_2<\textstyle\frac{1}{4}$ and $\rho_2\rho_3>\textstyle\frac{1}{4}$.

On the other hand, $P^{(3)}(y)=\sum_{\{i_1,i_2,i_3\}\subseteq[k]}{a_{i_1i_2i_3} y_{i_1}y_{i_2}y_{i_3}},$ and $a_{i_1i_2i_3}=2^{-k}\sum_{y\in C}{y_{i_1}y_{i_2}y_{i_3}}$.

Suppose $\rho_2\approx\rho_3$. For $y\in G_{\phi_l}$, we have for each $\{i_1,i_2,i_3\}\subseteq[k]$,

\begin{equation*}
\begin{split}
&a_{i_1i_2i_3} \\
&=\textstyle\frac{1}{k\sqrt{k}2^k}({k\choose{\gamma_1 k}}(6\rho_1(\textstyle\frac{4}{3}\rho_1^2-1)+o_k(1))-{k\choose{\gamma_2 k}}(6\rho_2(\textstyle\frac{4}{3}\rho_2^2-1)+o_k(1)) \\
&+\textstyle{k\choose{\gamma_3 k}}(6\rho_3(\textstyle\frac{4}{3}\rho_3^2-
1)+o_k(1))) \\
&\approx\textstyle 6c(\rho_1(\textstyle\frac{4}{3}\rho_1^2-1)+o_k(1))\frac{1}{k^2}.
\end{split}
\end{equation*}
\noindent and

$$\sum_{\{i_1,i_2,i_3\}\subseteq[k]}{y_{i_1}y_{i_2}y_{i_3}}
=(\rho_l(\textstyle\frac{4}{3}\rho_l^2-1)+o_k(1))k\sqrt{k},$$

\noindent hence

$$P^{(3)}(y)\approx 6c(\rho_l(\textstyle\frac{4}{3}\rho_l^2-1)\rho_1(\textstyle\frac{4}{3}\rho_1^2-1)+o_k(1))\frac{1}{\sqrt{k}},$$

\noindent $c$ is an absolute positive.

When $k$ is sufficiently large, we can determine $\rho_l$'s satisfying $\rho_1\rho_2<\frac{1}{4}$, $\sqrt{\frac{1}{2}}<\rho_2\approx\rho_3<\frac{\sqrt{3}}{2}$, $\gamma_l k$'s are integers such that $\varphi_l$'s can be disguised by $\psi$ to a balanced pairwise independent distribution, and the tri-linear term of Fourier spectra of $C$ is at least $\iota=\Omega(\frac{1}{\sqrt{k}})$.

\end{document}